\documentclass[mathleft
]{an}
\usepackage{graphicx}
\usepackage{times}
\begin{document}

\Pagespan{178}{181}
\Yearpublication{2008}%
\Yearsubmission{2007}%
\Month{11}%
\Volume{329}%
\Issue{2}%

\def \xmm {\hbox{\it XMM-Newton~}}
\def \chandra {\hbox{\it Chandra~}}

\title{Supernova remnants, planetary nebulae and superbubbles: \\
prospects for new \xmm observations}

\author{A. Decourchelle\inst{1}\fnmsep\thanks{Corresponding author:
  \email{adecourchelle@cea.fr}\newline}
}
\titlerunning{Instructions for authors}
\authorrunning{A. Decourchelle}
\institute{
Laboratoire AIM, CEA/DSM - CNRS - Universit\'e Paris Diderot, DAPNIA/Service d'Astrophysique, B\^at. 709, CEA-Saclay, F-91191 Gif-sur- Yvette C\'edex, France
}

\received{2007 Nov 1}
\accepted{2007 Nov 29}
\publonline{2008 Feb 12}


\abstract{Important results achieved over the last years on supernova remnants, planetary nebulae and superbubbles are briefly reviewed in the context  of X-ray observations. I intend to review the important open scientific questions in these fields, and the specific contributions that can be made by {\it XMM-Newton}.}

\maketitle

\section{Introduction}
The hot phase of the interstellar medium (ISM), known \, since more than 30 years, is generated by supernova (SN) explosions, and to a lesser extent by powerful winds from the progenitor stars (McCray \& Snow 1979; Spitzer 1990; Ferri\`ere 2001). 
Both act qualitatively in a similar way as they enhance the ISM metallicity through injection of mass enriched in heavy elements by the thermonuclear reactions taking place in stars and supernovae. They forge the structure of the ISM through injection of energy and are largely responsible for its multi-phase nature and turbulent state. 
Last but not least, supernova remnants are likely the birth place of Galactic Cosmic Rays and source of amplification of the magnetic field through particle acceleration at their shocks (Bell 1978; Blandford \& Ostriker 1978). 

X-ray observations are crucial to study these sources that shape the large view of galaxies: supernovae and superbubbles, stellar winds and to some extent planetary nebulae.

\section{Planetary Nebulae}

Planetary Nebulae (PNe) have been discovered  more than 200 years ago. However, the understanding of their origin, structures and evolution is just beginning with a better coverage of the entire electromagnetic spectrum (Kwok 2005). 
In particular, the nature and origin of their extended, spatially resolved X-ray emission is under close investigation since the launch of \chandra and \xmm  X-ray observatories (Akashi et al. 2007 and references therein). 
PNe are formed by the interaction of a tenuous fast wind with a copious slow Asymptotic Giant Branch (AGB) wind, that produces either elliptical or bipolar PNe (Stute \& Sahai \, 2006; Frank \& Mellema 1994). 
This interaction gives rise to \,\,\, shocks that heat the fast wind material to temperature above $10^6$~K. X-ray observations provide thus a way to shade light on the physical properties of the fast wind (speed, mass flux, opening angle, composition), and shaping of PNe. In the following, I will discuss some of these issues.

X-ray observations of PNe do show a relatively low temperature of the emitting plasma of the order of $(1-3)\times10^6$~K, while the observed velocity of the fast winds 
($\ge 1000$~km/s) implies a post-shock temperature greater than $10^7$~K. Heat conduction (or mixing) at the interface with the cool slow wind material may lower the temperature and increase the density of the fast wind. Other possibilities include a rapid weakening of the fast wind in terms of mass loss,  a slower moderate velocity wind of $\simeq 500$~km/s (post-AGB wind), or two opposite jets (collimated fast wind) as discussed in Akashi et al. (2007). 

Using \xmm observations, Gruendl et al. (2006) 
have shown that the spectrum of the planetary nebula NGC 7026 was best fitted with a model with nebular abundances than with one with stellar wind abundances. This result, in favor of a significant mixing with the cool nebular material, would provide an explanation for the observed low temperature. However, the statistics is too limited ($< 200$ photons) to get a firm conclusion on the composition.

Fast winds are produced during the late evolutionary stage of the central star. Through their mass loss, they eject late phase products (C, N, O and Ne) synthesized in the CNO cycle and He burning phase. X-ray observations of the shocked wind give thus access to these late phase products. Recently, in the brightest X-ray planetary nebula BD + 30 3639 observed with {\it Suzaku}, Murashima et al. (2006) 
found an extreme C/O and Ne/O enhancement over the solar values by a factor 30 and 5, respectively. This indicates that the X-ray gas is composed of helium-shell-burning products.

The X-ray luminosity of PNe is observed to be much lower than theoretically expected. While there are about \, 2000 Galactic PNe (Parker et al. 2003),  
less than 20 were identified as X-ray sources. They are essentially the young elliptical and bipolar PNe, but not the evolved ones. Only three bipolar PNe have been detected in X-rays, despite the observation of five other bipolar PNe with {\it Chandra}.
With {\it XMM-Newton}, the prospects are on one hand to increase the statistics on the X-ray spectrum of well identified PNe to provide stringent constraints on the nature and origin of the emission through their composition. On the other hand, we need to increase the number of observed PNe in X-rays to investigate the statistical properties of the central wind and collimated fast wind in elliptical and bipolar PNe. 
In the later case, observations should concentrate on the objects with closed bipolar lobes to confine the hot gas (Gruendl et al. 2006).
A relevant observing program on a few selected PNe with \xmm will require typically 1 million second observation to achieve significant spectral constraints.

\section{Supernova remnants}

In young supernova remnants (SNRs), the main physical process is the interaction of high velocity ejected material with the ambient medium. This gives rise to high Mach number shocks that heat the shocked ejecta and shocked ambient medium to X-ray temperatures, and accelerate particles to TeV energies. These sources are powerful X-ray emitters. Thermal emission of the shocked ejecta at $\simeq 1$~keV usually dominates the overall spectrum while synchrotron emission essentially located at the shock dominates the 4-6 keV continuum. This is well exemplified by the deep {\it Chandra} observation of Tycho (Warren et al. 2005).

\subsection{Nucleosynthesis products}

X-ray spatially resolved spectroscopy with {\it XMM-Newton} and {\it Chandra} has allowed to map the distribution of the shocked synthesized elements in the ejecta of young SNRs. While Tycho's SNR is the prototype of a type Ia SN, Kepler's SNR is probably a type Ia as well, but with substantial circumstellar material (Reynolds et al. 2007).  
In both remnants, a radial onion elemental structure is roughly preserved with iron toward the inner region and silicon in the outer ones (Decourchelle et al. 2001; Cassam-Chena{\"\i} et al. 2004a). 
Conversely, in the core-collapse supernova Cas A, there is an important spatial inversion of a significant portion of the SN core (Hughes et al. 2000a). 
A million second {\it Chandra} observation of Cas A has provided evidence of the bipolar structure of the Si-rich ejecta, and in particular of the counterpart of the northeast jet (Hwang et al. 2004). 
While \chandra is perfectly suited to study such small angular size remnants ($< 10$ arcmin), \xmm has the unique ability to provide detailed spatially resolved spectroscopy of the closer or more evolved SNRs. These remnants, like Puppis A, Cygnus Loop and Vela, do show enhanced metal abundances from the ejecta (see for example Katsuda \& Tsunemi 2006). They are important to study in order to understand the evolution of SNRs in terms of hydrodynamics, ISM enrichment and also particle acceleration. 

\subsection{Shock physics: temperature equilibration}

In young SNRs, the high Mach number shock propagates usually in a low density ambient medium ($\le 1\, \mathrm{cm}^{-3}$). The partitioning of the energy between the different species (e\-lectrons, protons,..) is not dominated by Coulomb collisions but by collective plasma processes, which are theoretically and observationally not well known. The analysis of Bal\-mer-dominated optical spectra from non-radiative SNRs \, allows to measure the ratio of the electron to proton tem\-perature (Ghavamian, Laming \& Rakowski 2007). 
In X-rays, using the high resolution spectrum obtained with the RGS onboard \xmm of an oxygen-rich knot in SN 1006, Vink et al. (2003) 
have been able to measure for the first time the temperature of the shocked oxygen. The measured Doppler broadening of the OVII line corresponds to an oxygen temperature of $\simeq 530 \pm150$~keV while the electron temperature is of $\simeq 1.5$~keV. This shows that the degree of temperature equilibration between the electron and ion is very small ($\le 5 \%$).

The high resolution spectroscopy is also a strong stren\-gth of \xmm for measuring the ionization conditions, the abundances and the temperature of other species than electrons, like ionized oxygen (and possibly other \, ions). This provides a very strong diagnostic on the partition of the shock energy into different species and is very relevant for understanding collision-less processes at the shock wave.

\subsection{Electron acceleration}

\begin{figure}
\begin{center}
\includegraphics[width=60mm,height=60mm]{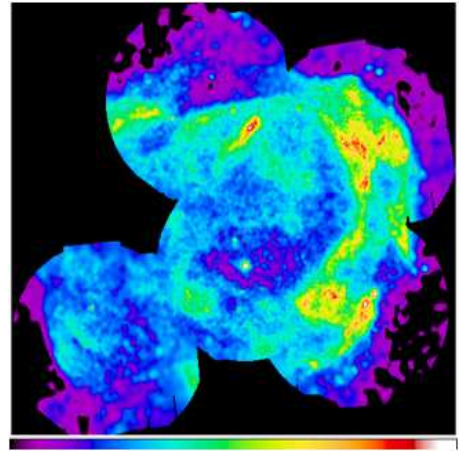}
\caption{\xmm 2-10 keV image of the synchrotron-dominated SNR G347.3-0.5 (or RX J 1713.7-3946).}
\label{label1}
\end{center}
\vspace{-0.2cm}
\end{figure}

SNRs are expected to provide the pool of Galactic cosmic rays up to the knee (at $\simeq 3\times10^{15}$~eV). Electrons are only a few $\%$ of cosmic rays, but they carry a lot of information on the mechanism of diffusive shock acceleration. While $10\,\%$ of the kinetic energy of SNRs is sufficient to account for the cosmic-ray energetic, a crucial requirement for the SNR scenario is to be able to accelerate particles at least up to the knee. It is where X-ray synchrotron observations can play a central role. Indeed, the maximum energy of accelerated electrons can be obtained through the measurement of the cut-off frequency of the synchrotron emission, which falls typically in the X-ray range. If the magnetic field $B$ is known, this measurement determines the maximum energy of accelerated protons. This has been done in a number of SNRs, and provides a typical maximum energy of the order of $80$~TeV (if $B \simeq 10\,\mu$G, Reynolds and Keohane 1999).

In the type Ia SN 1006, it has been possible to measure the azimuthal variation of the maximum energy along the SNR shock. Using {\it XMM-Newton}, Rothenflug et al. (2004) 
have shown very strong azimuthal variations that cannot be explained by variations of the magnetic compression alone. The maximum energy of the accelerated particles is higher at the bright synchrotron limbs than elsewhere. If the magnetic field is amplified at the limbs, the maximum energy of protons may be as large as $1000$~TeV there, while outside the limbs it can be around $25$~TeV for a compressed magnetic field of $\simeq 10\,\mu$G. The geometry of the synchrotron emission favors cosmic-ray acceleration where the magnetic field was originally parallel to the shock speed i.e. in polar caps rather than in an equatorial plane.

The morphology of the X-ray synchrotron emission provides also constraints on the intensity and configuration of the post-shock magnetic field. In most young SNRs, the X-ray emission just behind the blast wave has been observed to be spectrally featureless and confined in thin sheets (Hwang et al. 2002; Cassam-Chena{\"\i} et al. 2004a; Bamba et al. 2005). 
Two interpretations have been proposed to account for this filamentary emission: either the magnetic field is large e\-nough ($\simeq 100\,\mu$G) to induce strong radiative losses in the high energy electrons (Vink and Laming 2003), 
or the magnetic field is damped at the shock (Pohl, Yan \& Lazarian 2005). 
Both models predict different synchrotron morphologies \, and spectral shapes in X-rays (Cassam-Chena{\"\i}  et al. 2007). 

{\it XMM-Newton} is ideally suited to investigate particle acceleration through deep spatially resolved spectroscopy of the synchrotron dominated shell SNRs, which are all well extended (SN 1006, G347.3-0.5, Vela Junior). Such observations are required to measure the maximum energy of accelerated particles and its variation along the shock  as well as to provide estimate of the density all around the SNR. 

\subsection{Proton acceleration}

While electrons are expected to be accelerated like protons and to convey relevant information on the particle spectrum and maximum energy, they are only test-particle for the hydrodynamics. This is the protons that carry most of the energy of the accelerated particles. 

While we do observe in X-rays electrons accelerated at TeV energies in SNRs, a direct unequivocal evidence for ion acceleration is still missing. There are unresolved questions on the efficiency of cosmic-ray acceleration and on the fraction of the shock energy that can be tapped by the cosmic rays. Two possible complementary ways to answer these questions are discussed. 

One way is to observe the back-reaction of the accelerated protons on the shock structure and on the overall hydrodynamics of the interaction region. For an efficient ion injection, a large fraction of the energy goes into accelerating particles. As a consequence, the shock compression ratio gets larger (above the standard test-particle value of 4 for strong shocks) and the post-shock temperature gets lower. This is due to a change of the adiabatic index of the gas and to the fact that the highest energy particles can escape the shocked region of the SNR. This modifies the overall hydrodynamics of the interaction region, which becomes narrower (Decourchelle, Ellison \& Ballet 2000; Ellison, Decourchelle \& Ballet 2004). 
2-D simulations have confirmed these results (Blondin \& Ellison 2001). 
These predictions are confirmed by X-ray observations of some young SNRs (Hughes, Rakowski \& Decourchelle 2000; Decourchelle 2005; Warren et al. 2005). 

The other way is to observe the spectral signature of neutral pion decay in the GeV-TeV $\gamma$-ray regime (e.g. with {\it GLAST}), due to the collision of accelerated protons with those of the interstellar medium. At these energies, there is also a contribution of the Inverse Compton emission, which needs to be determined for a precise estimate of the pion decay emission. The knowledge of the population of TeV electrons, observable in X-rays, is crucial for this task as well as the knowledge of the optical and infrared photon environment around the SNR. The remnant G347.3-0.5 is an excellent illustration of the complementarity between X-ray and TeV $\gamma$-ray observations. This remnant, whose X-ray emission is entirely synchrotron, exhibits a very similar morphology in X-rays (Fig. 1, Cassam-Chena{\"\i} et al. 2004b) 
and in TeV $\gamma$-rays (Aharonian et al. 2006). 


\subsection{Prospects for SNRs with \xmm}

All the above scientific issues require {\it XMM-Newton} observations. Thanks to its high sensitivity, good spatial resolution and large field of view, there are a large number of projects on SNRs, that can be only carried out with {\it XMM-Newton}. Its large field of view allows to observe extended SNRs, which have not been the focus of current study as they require substantial amounts of exposure time for a relevant coverage of the object. They are however crucial for understanding various astrophysical issues like SNR evolution, mixing of the nucleosynthesis products with the ISM, acceleration of particles and energy partitioning at the \, \, shock. 
These are programs requiring typically millions second observations to have a representative high sensitivity coverage of these extended remnants. The above list of issues is not exhaustive. They are other important studies that require specifically {\it XMM-Newton}, like those related to the interaction of SNRs with interstellar clouds and to the statistical properties of galactic SNRs.

\section{Superbubbles}

The majority of O and B stars are grouped in association and $60\,\%$ of all core collapse supernovae are clustered. Superbubbles are engendered by the collective effects of supernovae ($\simeq 30$ in average) and stellar winds (Ferri\`ere 2001). 

There are a number of open issues in the field of superbubbles. What is the origin of the bright X-ray regions in superbubbles? Are they regions of thermally evaporated material or superbubble shells shocked by recent SNe? Why the amount of energy currently present in superbubbles is significantly less than the energy input expected from the enclosed massive stars over their lifetime (Cooper et al. 2004)? 
What is the origin of the nonthermal emission observed in some superbubbles (Smith \& Wang 2004)? 
Are superbubbles a relevant source for the acceleration of galactic cosmic rays (Parizot et al. 2004)? 
 
 $^{30}$Dor C is a good example of a well known massive star-forming region located in the $^{30}$Doradus region in the Large Magellanic Cloud (LMC), which was observed as the first-light image for \xmm (Dennerl et al. 2001). 
Deeper observations of \xmm have been used to obtain spectra of the $^{30}$Dor C superbubble. The hot gas is enriched in $\alpha$-elements (O, Ne, Mg, Si, S, Ar, Ca) by a factor of about 3 while the group of Fe-like elements is consistent with the LMC abundances. The additional $\simeq 5$~M$_\odot$ of oxygen in the hot gas require 2-3 high mass ($\ge 20$~M$_\odot$) recent core-collapse supernovae (Smith \& Wang 2004). 
Nonthermal emission has been detected as well in this superbubble (Bamba et al. 2004). 
However, the origin of this emission is still an open question: synchrotron from TeV accelerated electrons, Inverse Compton or nonthermal bremsstrahlung (Smith \& Wang 2004)?

The LMC is an ideal laboratory for observing superbubbles in X-rays, with a known distance and a faint interstellar absorption. About 20 superbubbles have been identified. The prospects with \xmm are to get deep exposures on specific superbubbles for which we can characterize the abundances of the thermal hot gas and the spectral characteristics of the observed nonthermal component. The sensitivity of \xmm is required for such study, as well as the large field of view which allows to observe a large number of superbubbles.



\end{document}